\documentclass[sn-mathphys-num,a4paper]{sn-jnl}
\usepackage{graphicx}
\usepackage{amssymb}
\usepackage{amsmath}
\usepackage{mdframed}
\usepackage{helvet}
\usepackage[misc]{ifsym}
\usepackage[leftcaption]{sidecap}

\sidecaptionvpos{figure}{t}

\newtheorem{definition}{Definition}

\def\abstractheadfont{\normalsize\bfseries\selectfont}%
\def\abstractfont{\normalsize\leftskip=24pt\rightskip=24pt\parfillskip=0pt plus 1fil}%
\def\keywordfont{\normalsize\leftskip=24pt\rightskip=24pt plus0.5fill}%

\makeatletter
\def\Titlefont{\reset@font\fontsize{14bp}{16bp}\sffamily\bfseries\selectfont}%
\def\Authorfont{\reset@font\fontsize{12bp}{14.5bp}\sffamily\bfseries\selectfont\boldmath}%
\def\bibfont{\reset@font\fontfamily{\rmdefault}\fontsize{9bp}{11bp}\selectfont}%
\def\sectionfont{\reset@font\fontfamily{\rmdefault}\fontsize{12bp}{14bp}\sffamily\bfseries\selectfont\raggedright\boldmath}%
\def\subsectionfont{\reset@font\fontfamily{\rmdefault}\fontsize{11bp}{13bp}\sffamily\bfseries\selectfont\raggedright\boldmath}%
\def\bmheadfont{\reset@font\fontfamily{\rmdefault}\fontsize{9bp}{11bp}\sffamily\bfseries\selectfont\raggedright\boldmath}%
\makeatother

\begin{document}

\title[Article Title]{Does Quantum Information Require Additional Structure?}

\author[]{\fnm{Ryszard} \sur{Horodecki\textsuperscript{1,2}}
\footnotetext{\Letter\hspace{0.45em}
Ryszard Horodecki \\
\hphantom{1}\hspace{1em}
ryszard.horodecki@ug.edu.pl
\vspace{1em}}
\footnotetext{\textsuperscript{1}\hspace{1em}
International Centre for Theory of Quantum Technologies (ICTQT), \\
\hphantom{\textsuperscript{1}}\hspace{1em}
University of Gdańsk, Jana Bażyńskiego 8, 80‑309 Gdańsk, Poland}
\footnotetext{\textsuperscript{2}\hspace{1em}
National Quantum Information Centre in Gdańsk, Jana Bażyńskiego 8, \\
\hphantom{\textsuperscript{2}}\hspace{1em}
80‑309 Gdańsk, Poland}}


\abstract{
\textnormal{We consider the status of quantum information in the quantum theory and based on the correspondence principle, we propose an interpretation of the wave function as a mathematical representation of quantum information. We consider Clauser’s analysis of incompatibility formulations of quantum theory in laboratory space and configuration-space in the context of local realism. Then, we introduce the hypothesis of quantum space of directly unobserved relations, which precede quantum  correlations, and are compatible with the Reichenbach common cause principle. The possible implications of the hypothesis are discussed in the context of the latest experimental and theoretical results on the dynamics of entanglement formation in helium atoms. Finally, we present the Chyliński model as an example of quantum relational continuum space, which predicts potentially measurable effects for the bound states.}}

\maketitle

\begin{flushright}
\emph{``Heisenberg’s motivation for studying physics 
was not only to solve particular\linebreak
problems but also to illuminate the discussion of broad philosophical question''}

\emph{Frank Wilczek
``Four Big Questions with Pretty Good Answers''}
\end{flushright}

\vspace{1cm}

\section{Introduction}

Quantum theory is the best theory we have to describe physical reality. ``All our former experience with the application of quantum theory seems to say: what is predicted by quantum formalism must occur in laboratory'' \cite{HHHH}. It can be said that it works very well, which is why some physicists claim that ``Quantum theory needs no interpretation'' \cite{Peres}. Nevertheless, there is a significant group of scientists for whom the lack of a coherent, commonly accepted interpretation is a source of discomfort. 
There have been many significant attempts to build a consistent interpretation, such as Copenhagen interpretation, Hidden Variable interpretations \cite{Bohm, Durr}, and Relativistic interpretations \cite{Everett, Fuchs_RMP, Rovelli}. However, there is still no consensus on what quantum mechanics is. There have been various propositions: “only on relative facts” \cite{Rovelli}, “only on correlations and only correlations” \cite{Mermin}, “on the only fields” \cite{Fayngold}, “All is $\psi$” \cite{Vaidman}, “The everything-is-a-quantum-wave” \cite{Vedral}.  
One of the obstacles to a deeper understanding of quantum theory is that it radically deviates from classical patterns. Even pioneer quantum mechanics formulated
 nearly a hundred years ago, it appeared as a kind of quantum inscription encoded in the language of Hilbert space - a set of rules specifying how to make probabilistic predictions of results of future measurements in laboratories. The scale of difficulty in its deciphering reflects the fact that the discoveries of such phenomena as EPR correlations (entanglement), quantum computing \cite{Deutsch, Feynman}, quantum teleportation \cite{Bennett_6}, existence of the quantum correlations beyond entanglement \cite{Ollivier, Henderson_Vedral, OHH} took place long after its birth. Quite surprisingly despite winning the battle with nature in an experimental demonstration of breaking Bell's inequality (Nobel Prize in physics 2022 \cite{Aspect}) ``...there is still no consensus about what such violations actually imply'' \cite{Raedt} (and ref. 
 therein).
Moreover, quantum theory does not offer any simple footbridge from the quantum world to our actual world (see in this context \cite{Landsman, Zwolak, RH_Korbicz, Paweł-Brandao-NC, Le}. This issue covers many aspects: it has a long history and a vast literature, including the famous Wigner paradox \cite{Renner, Żukowski}.

Paradoxically, although the concept of quantum information lies at the heart of quantum theory, it has rather little impact on its interpretative problems. The vast majority of considerations about the meaning of quantum formalism do not take into account the problem of the relationship between physical reality and mathematical models, including the role of universal constants.
This work does not aim to solve these drawbacks of quantum theory but rather attempts to consider the quantum inscription in light of the correspondence 
principle and the concept of quantum information.
In Sec.2, we introduce the correspondence principle between physical reality and mathematical models
and we discuss its significance in relation to physical models (theories) in the context of their classification under the roles of universal constants.
 In Sec.3, we briefly discuss the relativistic classical model and the standard (nonrelativistic) quantum model, emphasizing the roles of universal constants and symmetry groups. In Sec.4, we consider the status of quantum information in the standard quantum model, and based on the correspondence principle, we propose an interpretation of the wave function as a mathematical representation of quantum information. In Sec.5, we consider Clauser's analysis of incompatibility formulations of quantum theory in laboratory space and configuration-space in the context of local realism. In Sec.6, we hypothesize that quantum information requires the existence of an intrinsic relational quantum space in which reversible pre-measurements take place and which precedes the space of correlations actualized in Minkowski space. In Sec.7, we present Chyliński's model of quantum relational continuum space, which leads to measurable effects and, at the level of operational observation, is consistent with special relativity. 

\section{Heisenberg's Classification of Fundamental Models and the Correspondence Principle}

Looking at the impressive development of physical sciences, a fundamental
question constantly comes back to us: What is the reason why 
nature allows itself to be described by means of mathematical structures
and why, out of many potential structures, it accepts only certain selected ones, e.g., Riemannian geometry but non
set‑theoretic geometry with Banach‑Tarski paradox \cite{Banach}. This peculiar selectivity of
nature can be naturally related to the existence of an active observer equipped with
mathematical
tools, which organize his/her rational thinking about the world and allow
him/her to create models of physical reality. By the term active observer, we mean the conscious
observer endowed with free will who can have access to the system
directly or indirectly via control and measurement interactions
associated with an experimental setup. However, the action of an active observer is accompanied by the activity of nature itself due to the existence of fundamental interactions. Hence, the potential connection between the mathematical structure of models and the structure of the described reality is far from obvious \cite{Eckstein_P,Raedt_EPR,Raedt}. Two approaches can be distinguished \cite{Raedt_EPR}: the "theory-driven" approach, which assumes that the production of experimental data is governed by a pre-existing mathematical model, and the "data-driven" approach, according to which experimental data are considered to be immutable facts, which require to construct corresponding mathematical models. In this context, it is tempting to introduce a correspondence principle (CP): 

\vspace{1em}
\begin{mdframed}
Any consistent successful partial description of nature is a sort of  
correspondence between the structure of physical reality and the structure of its mathematical representation.
\end{mdframed}
\vspace{1em}

Here, by the term ``physical reality'', we mean a set of empirical observer-independence data, which includes all fundamental physical constants \cite{CODATA}.
The term ``partial description of nature'' expresses
the fact that, in practice, an observer has access only to a part of the world called a ``system''.  The term ``correspondence'' means that physical reality expressed in the language of specific mathematical structures is true according to the Aristotle’s theory of truth \cite{Stanford Encyclopedia of Philosophy, Aristotle}. The
problem is that the correspondence principle itself does not specify how elements of mathematical structure are related to 
physical reality. Therefore, the observer must somehow establish a set
of interpretative statements to extract meaning from 
mathematical formalism. He can always, in the spirit of Gödel, supplement
the model with additional structure before applying it.

It is significant that “canonical” models of reality, including the
Standard Model need constants parameters (fundamental constants) that
cannot be explained within the framework of the given model, and they
can only be determined from the experiment. Note that the correspondence principle allows for an ``operational'' interpretation of the model-independent
parameters, \emph{as
interaction constants between the mathematical structure of a model
and the structure of the reality it describes}.
These constants impose global realistic constraints on the
mathematical structure of models. These constants are necessary for the model  to reproduce the experimental results. In particular, they build
relationships between quantities that were thought to be
incommensurable.

Note that there is no characterization of fundamental constants, and
their list usually depends on what we accept as a fundamental model
(theory). Hence, not all constants are on the same footing.
Historically, the Planck’s  constants $h$, and speed of light $c$
were promoted to the role of universal constants. Here, we will use the
term ``universal constants'' for which we assume that they meet the
universality postulate: \emph{
values of universal constants measured in any reference system
always take the same values} (see in this context\cite{Weinberg,Stuckey,RH_comment}).

According to Heisenberg, the introduction of universal constants led to
abrupt changes in the paradigms of  four fundamental models (theories) \cite{Heisenberg}: i) relativistic classical (RC): $1/c \neq 0$ and
$\hbar = 0$, ii) the non-relativistic quantum (NRQ): $1/ c = 0$,
$\hbar\neq 0$ 
iii) the relativistic quantum field (RQF): $1/ c \neq 0$ and $\hbar\neq 0$.
This classification does not take into account the gravitational constant $G$, which may be of fundamental importance in the quantum-to-classical transition \cite{Private_Eckstein}.

Note that there were also attempts to build a quantum relativistic
models with $\hbar/c \neq 0$   \cite{Bethe_Salpeter, Ch_PRA, Klein}.  In the final part of the paper, we will present Chyliński's
relation-space-time model
\cite{Ch_PRA} which, as far as we know, has not been considered so far,
in any physical or philosophical context.

To our knowledge, there is no commonly accepted characterization of the basic models (theories). Here, we adopt a simple characterization that defines the underlying model as \empty{closed} and \empty{compact} \cite{Chyliński_książka}. We say that a given model (theory) is closed if the internal symmetry of its equations is the symmetry of its measurement space-time (empty background), and it is compact if it explains the structure of measures and clocks necessary to create its metric background.

\section{Relativistic Classical Model Versus Quantum Nonrelativistic model}

Obviously, every one of the above models is a kind of idealization, while its predictive power is determined by a specific paradigm intimately related to universal constants. Specifically, the RC model ($\frac{1}{c} \neq 0$, $\hbar = 0$) based on the Galileo-Einstein principle and the postulate that the speed of light in vacuum is the same for all observers leads to the concept of Minkowskian space-time $M$ and its Poincare symmetry group as a classical relativistic model of space-time. 

Note that according to CP, the mathematical structure of the canonic phase space can be directly attributed to the physical structure as the joint probability distributions assigned to the classical phase space regions unambiguously correspond to the possible states of the system in which it has specific properties. Thus, by knowing the state of the system we can say that we know the physical reality regardless of the measurement context. The later is only auxiliary tool for measurements that do not disturb the system. Thus information can be freely copied by repeatedly interacting with the system.
Clearly, the RC is not compact. As Einstein already noted: ``in the present state of theoretical physics, measures, and clocks must be considered as independent concepts, and we are far from a precise knowledge of theoretical foundations to be able to provide the structure of these objects.''

In contrast to the RC model, the role of CP in relation to the NRQ model is much more subtle, as the latter strongly deviates from classical patterns. It appears rather as a quantum inscription written in the language of Hilbert space where the observables represented by Hermitian’s operators and the quantum states by positive unit trace matrices cannot be directly attributed to the physical reality. Here, the measurement is active: It can change state and can also be part of quantum operation.  Therefore, in contrast to the RC model the measurement context cannot be ignored.
What most differs the NRQ model from RC one is quantum principle of superposition, which has quantum origin.

This step change of the paradigm in the description of nature has its origins in the fact that the mathematical structure of the NRQ includes Planck’s constant $\hbar \neq 0$ implying i) the ultimate limit on the non-commutativity of measurement operators that impose constraints on the possibility of extracting information by measurement from a quantum system\cite{Brukner, Walls}, so separation of the measured system from the measuring device ceases to be obvious as in the RC, ii) breaking the rescaling of the spatiotemporal extension of the systems by determining the finite volume of the phase cell, iii) the existence of the quantum information carriers (QIC).
The NRQ model has an amazing property; even though it does not include the speed of light $c$, it is consistent with the results of the RC model such 
as the impossibility of superluminal communication.
     
The NRQ model applies to the phenomenological regime in which energies are low, and the effects of special relativity and high-energy physics can be neglected. The relativistic effects can be described within the framework of the RQ, which is based on the Lorentz-Poincare symmetry $M_4$. Note, however, that the NRQ model is fundamental because it is closed and compact. Closed theories in Galilean $G_4$ space-time can introduce hypothetical particles characterized by their absolute masses and spins. In particular, the intrinsic symmetry of the non-relativistic Heisenberg’s, Schrödinger’s, and Pauli’s equations is also the symmetry of their empty Galilean group $G_4$ background. As one knows, the unitary representations of the $G_4$ have a non-trivial projective nature with physical consequences: Bargmann’s superselection mass rule \cite{Bargmann}, the noncommutativity of operators momentum and position, the possibility of a spin angular momentum (Pauli’s model). The ``elementary'' particles are defined by irreducible projective unitary representations of the Galilean group in some Hilbert
space, which do not interact directly but via a field that mediates the interaction between particles \cite{Bargmann,Newton_Wigner}. It is generally accepted that the NRQ is neither universal nor the best model of reality, but it is certainly fundamental, and it can be a starting point for more advanced models such as RQ or S-matrix formalism. It is compact in the sense that it reproduces the extent of stable macroscopic measures with a hierarchical structure (quarks, nucleons, nuclei, atoms, molecules\ldots). Since NRQ was completed, there have been times when physicists thought that its ``quantum code'' was deciphered completely, like the Rosetta Stone. Therefore, it was a big surprise when Einstein Podolski and Rosen\cite{EPR}
discovered that NRQ predicts the existence of ``spooky'' correlations between properties of particles regardless of distance called entanglement \cite{Schrödinger}. Even though more than half a century has passed since its discovery, its full physical meaning still remains far from clear. Recently, it was argued that ``\ldots{} a violation of Bell-type inequalities by experimental data merely reflects the properties of the process, chosen by the experimenter, to select groups of data, not an intrinsic property of the data themselves.''\cite{Raedt}. Quantum entanglement and Born law break fundamental features of the RC: the separability and classical determinism, but the full significance of these facts seems to require new concepts.

\section{Quantum Information Puzzle}

There have been many obstacles to deciphering the potential physical implications of rich structure NRQ. It was only a series of groundbreaking theoretical and experimental papers showed that at a fundamental level, nature reveals the existence of a new, subtle property of physical systems - quantum information with no classical analog \cite{Jozsa}. Quantum entanglement, the most non-classical feature of quantum information, manifests itself at both macro \cite{Toth} and the micro scale \cite{Nandi,Time Delays,Kutak}.
Entanglement can be measured, processed, manipulated, and stored \cite{Bennett, Vedral_2, Toth, HHHH, Zeilinger_RMP, Modi}. 
The discovery that quantum information is a new subtle resource for non-classical tasks
\cite{Bennett_Brassard,Ekert91,Deutsch_Jozsa,Shor,Grover,Paweł_aktywacja,Paweł-Ekert,Michał_Nature,Bourennane,Karol_Winter}
revealed its operational meaning, which was of great importance for the development
of quantum information theory and quantum technologies.

Although the notion of quantum information was known before \cite{Ingarden}, it was
only the concept of intact transmitting quantum states in the papers
of the teleportation \cite{Bennett_6} and Schumacher quantum coding \cite{Schumacher95} that was
revolutionary because, at that time, the extremely epistemic
Copenhagen's interpretation was dominant. According to it, the wave
function $\psi$ (quantum state) is merely the description of the
measuring and preparing apparatuses. From such a point of view, it was
rather impossible to think that the message could be something else than
Shannon’s classical information.
Since $\psi$ is a central concept in the structure of quantum formalism the correspondence principle \emph{allows the interpretation of the $\psi$ wave function as a  mathematical representation (image) of the
quantum information.}

Importantly, the above interpretation (called here correspondence interpretation) goes beyond the dilemma of the Scylla of 
ontology and the Charybdis of instrumentalism, and it is consistent with the well-known
fact that in quantum models, the fundamental symmetry constraints are imposed on
the quantum state rather than on probabilities.

To our knowledge, there no formal definition of quantum information. Intuitively, one can think that \emph{the quantum information is what is carried by
quantum systems.}
The idea that a concept of quantum information should be regarded as a fundamental ingredient in quantum physics has been proposed \cite{Jozsa,IBM} from a different points of view. In \cite{IBM} the argument was mainly based on the concept of quantum information isomorphism, according to which the quantum description of nature is isomorphic to nature itself. Here we extend this approach by the correspondence principle (CP), which goes beyond a purely mathematical context.

Although so far, there is no formal definition of quantum information, the mathematical structure of Hilbert space imposes strong constraints on processing
of quantum information \cite{Horodecki_QI}: \emph{There is no universal broadcasting machine capable of broadcasting arbitrary quantum states.} Note that in contrast to quantum information, its mathematical representation $\psi$ can be broadcast. Thus, we would say that quantum information does exist, yet it is not just the wave function, but it is represented by it.
Below, we present the examples that support this view against the narrow Copenhagen treatment of the wave function (from \cite{IBM}). A paradigmatic example is quantum cryptography. Namely, the key distribution schemes show that there is a feature that is naturally ascribed to the quantum information carriers (QIC). It is clear that there must be quantum communication between Alice and Bob to achieve secret key \cite{Bennett_Brassard}. Indeed, Alice cannot simply send a wave function $\psi$ written on paper to Bob, as Eve could read it out without disturbing it. However, by sending QIC, Alice and Bob can achieve something impossible within the ``classical'' world. Another example is the direct calculation of quantum state function $f(\varrho)$ \cite{Paweł-Ekert}. It can be viewed as a kind of quantum computing operating on purely quantum input. This approach is natural if the quantum state $\varrho$ is viewed as representing real quantity to be processed rather than state of our knowledge. 

The above examples show clearly that the quantum information exists physically, and it cannot be identified with its mathematical representation $\psi$.     
Curiously, quantum entanglement - a special form of quantum information is relative due to \emph{the relativity of the concept of physical system} \cite{Zanardi, Dugic, Dugic2, Dugic3} and it is a
specific feature of the composite system structure \cite{Vedral3, Caban_Podlaski,Dugic2}.
It depends on the choice of factorization
of the underlying composite Hilbert space.
 For example, if we decompose the Hilbert space $H$ of the composite system $S$ as a tensor product of smaller spaces $H = H_1 \otimes H_2, \ldots, \otimes H_N$ then the unit vectors and the states of the system can be very different, so different decompositions can have different properties. This difference is due to the fact that the symmetry group $U(d_1) \times \ldots \times U(d_k)$, $d_i = \dim(H_i)$ is smaller than $U(d)$, $d = \prod^k_{i=1} d_i$.
In connection with this, the concept of quantum reference frame was proposed according to which ``a particular division of a composite system into subsystems constitutes a kind of reference frame called a meronomic reference frame \cite{Hulse_Schumacher}. In particular, it has been shown that in one quantum meronomic frame, separable (product) states may become entangled and vice versa \cite{Hulse_Schumacher,Czachor_N}.

In should be noted here that the relative character of quantum entanglement appears also when quantum mechanics is combined with special relativity. In particular recently it has be demonstrated that the existence of the entanglement encoded in the wave function is not universally agreed upon by different moving observers \cite{Nagele,Caban}. This, in particular, leads to ambiguity in interpreting the entanglement as a physical resource and in connection with this it was suggested that 
 a true resource should be quantified in a basis-independent way \cite{Nagele}.

Note that in the case for quantized electromagnetic fields, the “quantum” and “classical” features of
quantum information depend on the preparation of the initial states and the method of
detection \cite{Banaszek}. Namely, in contrast to classical communication, where the information
is encoded in classical properties of electromagnetic field (intensity or phase) quantum
communication requires to send quantum information carriers, such as nonclassical states
of light, e.g., squeezed states or Fock states that carry a well-defined number of photons.
In the typical communications links, the task is an optimization overall ensembles of
input quantum states used to carry information under relevant physical constraints.
It is the Holevo theorem that provides a tight bound on the mutual information
attainable for a given ensemble of input quantum states.

\section{Laboratory Space Versus Configuration-Space and Local Realism}

It still remains unclear why the NRQ model predicts the existence of stable quantum structures and EPR correlations and, at the same time, is consistent with RQF. 
Note that to protect the latter from superluminal speeds, the a priori
postulate of local commutativity is necessary \cite{Hagg}. However, the RQF is not strictly limited by the speed of light
\cite{Peskin}. In this context, the  existence of the 
EPR correlations naturally lead to the question of their physical nature and
still is subject to a wide debate \cite{Di Biagio, {Raedt_EPR}}. 
Interestingly, Gisin suggested ``that quantum correlations somehow
arise from outside space-time''\cite{Gisin_QT}. Clauser and Shimony had
previously suggested that ``Either we must completely abandon the
realistic philosophy of almost all practicing physicists or
fundamentally revise our concept of space-time'' \cite{Clauser_Shimony}. The second part of
the above alternative is a big challenge since Lorenz's symmetry 
imposed by heavy devices is an \emph{a priori}
measurement condition of any physical model \cite{Bohr}. On the other hand,
this challenge encounters an intriguing problem of
incompatibility formulations of quantum mechanics in laboratory space and
configuration-space, which has recently been the subject of deep
consideration in the context of local realism and correlation EPR
\cite{Clauser}. Here, laboratory (lab) space ``is the three-dimensional
space in which we live, and the space in the Euclidean geometry
is understood''\cite{Clauser}. Every point $\mathbf{r}_{lab}$ within lab has a unique position in what is called lab space.
    
According to Born's original formulation, the wave function
$\psi_{lab}(\mathbf{r}_{lab}, t)$
represents a field that propagates in lab space and correspondingly
the real-valued probability density 
$|\psi_{lab}(\mathbf{r}_{lab}, t)|^2$
also propagates as a classical field. Thus, Born's modification of
the interpretation of the wave function is expressed only in its
statistical interpretation. Moreover, lab-space
formulation applies only to the single particle
systems, and there is no strict method to extend this formulation  to $n \geq 2$
particle systems. In this context, Clauser notes ``This
impossibility might be expected since Local Realism’s formulation
and Bell’s Theorem forbids it. Correspondingly, experiments that
refute Local Realism also refute a lab space formulation of quantum
mechanics.''

Von Neumann introduced the abstract configuration-space formulation of
quantum mechanics for the system consisting of $N$ particles, which
uses a very general abstract $k$-dimensional argument space for the
wave function ($k=3N$), which
enables a system with $k$-degrees of freedom: $\mathbf{r}_{1}, \ldots, \mathbf{r}_{N}$.
A configuration space wave function is defined as a complex function defined in the configuration space
$\mathbb{R}^{3N}$.
\[ 
\psi_{conf,N}: \mathbb{R}^{3N} = \mathbb{R}^3 \times \cdots \times \mathbb{R}^3 \to \mathbb{C}, \quad
\psi_{conf,N}(\vec{\mathbf{r}}_1, \ldots, \vec{\mathbf{r}}_N) \in \mathbb{C}
\]
Clauser argues that the quantum configuration space formalism far surpasses the lab space formalism 
which is limited to describing only a single particle system and he concludes that quantum mechanics requires an explicitly unambiguous configuration-space formulation.
On the other hand, he notices:
``if we assume, according to the Copenhagen
Interpretation of quantum mechanics that abstract configuration-space is inherently impenetrable, then it allows for magical action
at a distance, and no objects are necessary to exist as things in the
laboratory space'' \cite{Clauser}.

\section{Quantum Relational Space Hypothesis}

In this section, we introduce the hypothesis of existence of quantum relational space (QRS). We start with the observation that, in accordance with
the correspondence interpretation of the wave function we can claim that the configuration wave function $\psi_{conf}$ is a mathematical representation (image) of quantum information. Consequently, the latter must have a non-spatial nature. Unfortunately the $\psi_{conf}$ 
space formalism itself does not provide any conceptual model to understand
its ``inner workings'', although intuitively, it seems to hide a deeper level of reality. Despite the vast knowledge about entanglement, it is not still
clear where it has its habitat and how it ``coexists'' with fundamental
Galilean and Lorentz-Poincare symmetries.
In 1999, Gisin’s team conducted a Michelson-Morley-like experiment which
showed that the 2-photon interferences were still visible,
independently of the relative velocity between Alice's and Bob’s
reference frames (actually, the magnitude of the velocity was constant
at $100 m/s$, but its orientation varied) \cite{Gisin_Sunday}.

Gisin's experiment suggests that the von
Neumann's collapse of the quantized electromagnetic field cannot be viewed as a real
causal process and quantization of the field definitely renders it no longer describable
as “real stuff in real space-time”, \cite{Clauser,Clauser_Gdansk}. In a result, Clauser concludes that  
''entanglement is not allowed in lab space'' \cite{Clauser_Gdansk}.
While he points out that the quantum configuration-space formalism captures $N$-particle entanglement. Intuitively, the multipartite nature of entanglement implies the existence of associated internal relational structure \cite{Esfeld}. 

Note that the notion of relation is more general than the notion of interaction. Remarkably, the entanglement structure encodes quantum relations that can exist even when there are no interactions, which implies non-local Bell violating correlations. It suggests that the nature of quantum information is, in principle, relational. Therefore, it seems reasonable to adopt the QRS hypothesis: 

\vspace{1em}
\begin{mdframed}
The fundamental physical interactions between quantum objects generate directly
unobservable relations, which precede quantum correlations, and are the source of physical properties of quantum objects.
\end{mdframed}
\vspace{1em}

Here we assume that every completely isolated system can be
associated with an intrinsic quantum space $Q_{rel}$ 
of directly unobservable relations generated by unitary interactions.
We assume that the space of quantum relations involves intrinsic quantum degrees of freedom such as spin, color, and smell, and it precedes the space of directly observable correlations $Q_{cor}$,
which is some set of events of the Minkowski $M_4$ space-time (Fig.1).

\begin{SCfigure}
  \includegraphics[width=0.6\columnwidth]{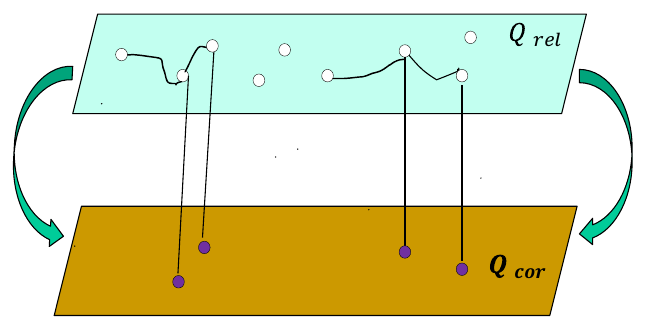}
  \label{fig1}
  \caption{$Q_{rel}$ relation space of “pre-measurements” precedes the space of correlation events $Q_{cor}$. Arrows indicate irreversible decoherence}
\end{SCfigure}

We are not able to determine the mathematical structure of the hypothetical relational space $Q_{rel}$. The simplest way to get around this difficulty is to assume that the relational space structure is 
isomorphic to the Hilbert space tensorial structure. Then, we can say that quantum models provide instructions on how to model a structure of $Q_{rel}$ in terms of Hilbert space formalism. 
In the multipartite dynamic scenario, building of quantum relations corresponds to the situation when the interaction (entangler) between systems belonging to partial Hilbert spaces $H_1$, $H_2$, 
$\ldots$, $H_N$, builds the total Hilbert space $H = H_1 \otimes H_2 \otimes \ldots \otimes H_N$ of the final entangled state. Recently, the experimental implementation of forming of entanglement between the ion and the photoelectron in a helium atom was demonstrated \cite{Nandi}. 
Quite lately, by the numerical solution \textit{ab initio} of the Schrödinger equation, it has been shown that 
building of directly unobservable quantum relations between light-field driven electronic dynamics in residual ion 
takes place in a short (attosecond) time before the photoelectron leaves the atom \cite{Time Delays}.
In particular, it have been shown that the delay of the photoelectrons carriers information on the entangled light-field dressed electronic dynamics in the residual ion.

Note that the
 algorithmic generation of genuine multipartite entangled states by coherent unitary deformation operations can be viewed as a kind of modeling of the
$Q_{rel}$ space \cite{Mahdavipour}.
In particular, the pure state $\psi$ of a composite quantum system encodes information
about all relations (amplitude and phase) between its subsystems, which may be related to
``the correlations without correlata'' \cite{Mermin}. 
Perhaps the most adequate representation of $Q_{rel}$ space is von Neumann's reversible pre-measurement scheme modeled by the unitary evolution coupling the measurement apparatus and the measurement system which involves information transfer caused by interaction. This interaction can be realized, in particular, by specific unitary Ozawa’s Hamiltonian evolution, which does not need a description of the measurement-induced state collapse, while it provides a cut between the measurement apparatus and the measured system \cite{Pranzini}.

Note, however, that the Hilbertian ``isomorphic constraint'' imposed on the  of $Q_{rel}$  may be too restrictive. In particular, it goes beyond Tsirelson's problem \cite{Gallego} and may rule out the description of potential ``exotic'' effects such as quantum gravity or ``stronger nonlocal correlations`` \cite{Gallego}. This may require more sophisticated mathematical structures such as non-separable Hilbert space \cite{Weaver}, quantum relations formalism \cite{Halvorson, Gallego}.

Interestingly, the model of quantum relational continuum space based on the configuration-space of relative coordinates and momenta, whose shapes are related to each other through unitary transformations, was proposed \cite{Ch_PRA} (see Sec. \ref{sec:Chylinski}). Note that there is no
reason to assume that the internal symmetry of the $Q_{rel}$ space must be the Lorentz symmetry.
Obviously, the assumption on the perfect
isolation of the system is a realistic idealization, as the unitary interactions do not yield outcomes \cite{Kastner2021}, thus
the system cannot ``recognize'' itself.
Hence, a full description of the pre-measurement does not imply a description of the output production.
It is intimately connected with the measurement
problem – ``Achilles' heel'' of quantum modeling of the world \cite{Renner, Żukowski}. For additive conserved quantities, the model of the measurement was proposed based on dilated Hilbert space that includes both the degrees of freedom of the measurement apparatus and the measured system \cite{Sewell, Sen}. 
In realistic scenarios, however, the observer is not able to control all the degrees of freedom environment of the system and broadcast of the results \cite{Zurek}. Thus, he is forced to use
a formalism that includes open systems and irreversible decoherence (i.e., \cite{Gorini, Lindblad}).
 In contrast to the view that every interaction is a measurement, 
 \cite{Rovelli}  (see of \cite{Brukner_noRowelli, Jacques L. Pienaar}) 
we distinguish here a class of projective, strong, and
instantaneous measurements that
are subject to Lorentz’s symmetry imposed by heavy bases as an a priori condition for the measurement of any physical model \cite{Bohr}. Such measurements will lead to an actualization of quantum relations, generating the correlation space $Q_{cor}$,
which can be viewed as a set of events of the Minkowski $M_4$ space-time.

In this sense the relation
space $Q_{rel}$ precedes $Q_{cor}$ a set of the actualizations (clicks) appearing as a result of irreversible 
measurements with specially prepared macro-objects (detectors) in
finite regions $C$ of $M_4$.
This is compatible with the dynamic conception of Minkowski space-time, which involves irreversible processes \cite{Weinert}
Then, it is reasonable to introduce measure $\mu(C)$ as the probability of a ‘click’ provided that a suitable detector operates in region C \cite{J_Geom_Phys}. 
This seems consistent with the suggestion \cite{Heller} that the quantum probabilistic structure is ``overbuilt`` over $M_4$, while the latter is only a ``carrier`` for the probabilistic measures. 
        
It should be noted that the
QRS hypothesis is compatible with Reichenbach common cause principle \cite{ Reichenbach} and Gisin’s conjecture that
the ``quantum correlations arise from outside space-time``.
Indeed, according to the QRS hypothesis, the quantum correlations have their origins in the  $Q_{rel}$ space. Interestingly recent results - experimental \cite{Nandi} and theoretical \cite{Time Delays} prove that quantum relations associated with the forming of the corresponding wave-packets do not emerge
instantaneously, but require a short, finite time.

Finally, if the QRS hypothesis is correct, it will shed new light on what is termed in the EPR folklore as "spooky correlations". Namely, in correlation measurements, the original quantum relations are destroyed  by strong measurements   interactions with the apparatus  and at the same time are actualized  as quantum correlations that ''remember'' their initial relational non-spatial feature of non-spacial origin. This peculiar feature makes quantum correlations to reveal as instantaneous in classical space-time. In this sense they are over cross-system and therefore they can be interpreted as observer-independent resource for quantum tasks (see of \cite{Nagele}).

\section{Chyliński's Quantum Relational Continuum Space}
\label{sec:Chylinski}

In this section, we present Chyliński's model of a relational continuum space
based on the configuration-space of relative coordinates. 
The NRQ model ($1/ c = 0$, $\hbar\neq0$) 
maintains the classical continuum of directly observable events 
$X^\mu=(\mathbf{X},X_0)$, $\mu=1,2,3,4$,
as an elementary background of most physical models. On the other hand, 
according to the uncertainty relations, the exact determination of
4-momentum $P^\mu=(\mathbf{P},P_0)$,
$P^\mu(\Delta P^c \to 0)$ 
rules out the $x$ localization of its quantum carrier in  
$M_4$ since $\Delta X^\mu \sim h/\Delta P^\mu \to \infty$.
It destroys in an uncontrollable way the ``quantum-potential'' structure $\psi$
of the isolated system under measurement. As one knows, the direct
p-measurements of the scattered particles in the asymptotic region do
not disturb the structure of $\psi$ 
state, which is consistent with the state of isolation of the
investigated system. The typical momentum measurements require
measuring position and inferring momentum, which, however, do not
violate the uncertainty principle. Note that there are also momentum
measurements that do not rely on measuring position based on the Doppler effect, the energy-sensitive single-photon detector\cite{Freericks}.

It is not obvious that Lorentz symmetry of “heavy bases” valid
micro-states when it is extrapolated to the interaction region. As
we know the quantum $p$-$x$ duality establishes one-to-one correspondence
between the p and x representations of a Hilbert vector $|G\rangle$ as
\cite{ChPolB}
\begin{eqnarray}
  \langle x | G \rangle &=& \int dp U(x,p) \langle p | G\rangle \nonumber \\
  \langle p | G \rangle &=& \int dx U^\star(p,x) \langle x | G\rangle
\end{eqnarray}

Note, however, that in this abstract formulation of the state $G$ 
of system $S_M$ the physical meaning of the $x$ and $p$
variables remain undetermined. According to the $M_4$ 
framework, the $p$'s denote four-momenta of the corresponding degree of freedom of $S_M$ 
represented in some reference frame $S$.
However, 4-symmetry revealed the experimental role of the asymptotic language $p$
of relativistic kinematics, which was enhanced by relativistic
$S$-matrix formalism. Thus, we can first measure the $S$ 
matrix elements (cross-sections) parametrized by Mandelsztam
$L$-invariant variable $s_k$
determined by asymptotic 4-momenta of free particles of the initial $|i\rangle$ and final $|f\rangle$
asymptotic states of the collision process
\begin{equation}
  \label{eq:4}
  S_{fi}= \delta^{(4)}(P_i - P_f)T_{fi}(s_1, \ldots, s_k), \quad s=(P_i - P_f)^2
\end{equation}
where $P_i$ and $P_f$ denote 4-momenta of the whole fully isolated system inside which the
quantum-potential collision takes place. Having cross-sections, we can
evaluate from Eq. \ref{eq:4} the $x$ representation of the corresponding structures, which is determined in
the $p$
language. In particular, the $S$-matrix formalism was used to estimate
(in a non-relativistic approximation) the scattering cross-sections
in collisions of electrons with atomic nuclei, based on an original
Chyliński \cite{Ch_PRA} model of internal quantum relation continuum $Q_{rel}$
space. To better understand the latter let us briefly recall the
singularity of Galilean space $G_4$. 
For this purpose, consider the simplest composite 2-body system $S_M=S_1 + S_2$ 
configuration-space $E_3(\mathbf{r}_1) \times E_3(\mathbf{r}_2)$
based on an arbitrary reference frame $S$
parametrizing  $G_4$. The $G$-absolute
time parametrizes evolution of all possible of $S_M$ 
generated by Hamiltonian 
\begin{equation}
  H^G = \mathbf{P}_1^2/2m_1 + \mathbf{P}_2^2/2m_2 + V((\mathbf{r}_2 - \mathbf{r}_1)^2)
\end{equation}
which depends on $S$
via kinetic energy term while  the potential $V$ is a $G$-absolute
relational shape. Using the well-known point and canonical transformations 
\begin{eqnarray}
  \mathbf{y}^G &=& \mathbf{r}_2^G - \mathbf{r}_1^G, \mathbf{R}^G= a\mathbf{r}_1^G+(1-a)\mathbf{r}_2^G; \; a=m_1/M \\
  \mathbf{q}^G &=& a \mathbf{p}^G - (1 - a)\mathbf{p}_1^G, \quad \mathbf{P}^G= \mathbf{p}_1^G + \mathbf{p}_2^G
\end{eqnarray}
where $\mathbf{p}^G$, $\mathbf{P}^G$ are the momenta canonically conjugate to $\mathbf{y}^G$, $\mathbf{R}^G$
respectively, we are passing to the new meronomic frame reference
for the system $S_M$ in which the internal ($\mathbf{y}^G$, $\mathbf{q}^G$)
and  external ($\mathbf{R}^G$, $\mathbf{P}^G$)
degree of freedoms respectively are dynamically separated. Note
that the $\mathbf{y}_G$ and $\mathbf{q}^G$
are invariants of the Galilean transformation between reference frame 
$S_M$, $S'_M$:
$\mathbf{y}^G=\mathbf{y}^{G'}$ and $\mathbf{q}^G=\mathbf{q}^{G'}$, hence they measure neither the localization of $S_M$
nor its velocity in the $G_4$.
In result  the total  operator Hamiltonian $\widehat{H}^G$ of $S_M$ of  
after Schrödinger quantization in  the $G_4$ space can be written in the form: 
\begin{equation}
  \widehat{H}^G=\hat{h}^G(\mathbf{y}^G, \mathbf{q}^G)+(\widehat{\mathbf{P}}^G)^2/2M
\end{equation}
where that the Galilean absolute internal Hamiltonian $\hat{h}^G$ 
becomes Galilean-form-invariant. Consequently, the energy eigenstates of the $S_M$
are product states 
\begin{equation}
  |\psi^G(\mathbf{R}^G; \mathbf{y}^G, \mathbf{q}^G) \rangle = 
  |\psi^G(\mathbf{R}^G) \rangle \otimes |\psi^G(\mathbf{y}^G, \mathbf{q}^G) \rangle
  \label{eq:8}
\end{equation}
which
are plane wave states for the center of mass and bound and unbound
states of the relative motion. However, these eigenstates are
entangled states of the two original subsystems of $S_M$ (see in this context \cite{Dugic2}).

In particular, the internal states of $S_M$ are the solution of the Schrödinger equation 
\begin{equation}
  i\hbar\frac{\partial}{\partial \tau^G} |\psi^G \rangle = \hat{h}^G | \psi^G \rangle
  \label{eq:9}
\end{equation}
where $\hat{h}^G$ in the internal Hamiltonian operator.

Due to the separability of the internal from the external degrees of freedom of $S_M$
it is convenient to introduce the internal relation continuum space the $\mathbf{Q}_{rel,4}^G$
composed of internal space $\mathbf{Q}_{rel,3}^G$
spanned by a relative coordinate $\mathbf{y}^G$
and the internal time $\tau^G$, which, up to the translation constant, can be identified with the absolute Newtonian time $t^G$.
Note that the symmetry of Eq. \ref{eq:9} consisting of rotations in $\mathbf{Q}_{rel,3}^G$
and translation in $\tau^G$
can be viewed as the symmetry of empty space $\mathbf{Q}_{rel,4}^G$.
Consequently all internal characteristic of $S_M$ in $\mathbf{Q}_{rel,3}^G$
are Galilean absolute. Moreover  $\mathbf{Q}_{rel,4}^G$ and $G_4$
deal with the same Galilean absolute intervals $r^G = |\mathbf{y}^G|$ and $\Delta r^G$,
in addition the points $(\mathbf{y}^G, \tau^G)$ of $\mathbf{Q}_{rel,4}^G$
are determined through the point (events) of $G_4$.
Thus the G-symmetry admits a relational absolute space $\mathbf{Q}_{rel,4}^G$ 
which is the Cartesian product $\mathbf{Q}_{rel,4}^G = \mathbf{Q}_{rel,3} \times T$,
which operates with two absolute intervals 
\begin{equation}
  r^G = |\mathbf{y}^G| \quad\textnormal{and}\quad \Delta \tau^G=\Delta t^G
  \label{eq:11}
\end{equation}
and it is isomorphic to the Galilean $G_4$ space-time since.
\begin{equation}
  \mathbf{y}^G = (\mathbf{r}_2^G - \mathbf{r}_1^G) |_{\Delta t^G = 0}, \quad \Delta \tau^G = \Delta t^G
  \label{eq:12}
\end{equation}

Thus, the peculiarity of geometry $G_4$
comes down to the fact that the Galilean ``eventism'' coexists with the ``relationism'' of 
$\mathbf{Q}_{rel,4}$.
The isomorphy of $\mathbf{Q}_{rel,4}$ 
to $G_4$ implies that $G_4$ gives room to both states and processes in the nonrelativistic limit  
$c \to \infty$.

\subsection{Internal Relation Quantum Space Time $\mathbf{Q}_{rel,4}$}

A hypothesis of the relation continuum $\mathbf{Q}_{rel,4}$
is closely connected with the isolation of physical systems, which
extends to finite universal constant $c$ the absolute nature of
Galilean relative coordinates and absolute Newtonian time. To extend
the absoluteness of $\mathbf{Q}_{rel,4}^G$, let us  consider first the  block  dual space-times 
$\widetilde{M}_4(x_\mu)$ and $\widetilde{M}_4(p_\mu)$  
where $x_\mu = X_\mu^{(2)} - X_\mu^{(1)}$ and $p_\mu=P_\mu^{(2)} - P_\mu^{(1)}$
are relative coordinates and 4-momenta, respectively, whose shapes
are related to each other by the unitary transformation $U$.
Note that the points of the  block (indivisible) space-time $\widetilde{M}_4(x_\mu)$
remain completely unlocalized in measurement space-time $M_4$

Now consider the directly measurable  observables $\widetilde{G}(p_\mu^2)=\widetilde{G'}(p_\mu'^2)$,
assuming that their $x$ representations are well defined, i.e., the corresponding integrals converge, in the asymptotic zone $L$-form-invariant function. Then the $M_4 - x$ representation of Hilbert vector $|G\rangle$ is given by 
\begin{equation}
  \langle x | G \rangle =
  (2\pi \hbar)^{-4}\int_{D} d^4 p \widetilde{G}(p_\mu^2) \exp [i (p_\mu x_\mu) / \hbar] 
  \label{eq:13}
\end{equation}
where $\xi_\mu$ are two arbitrary four points (events) in $M_4$ and $p_\mu$ 
are relative 4-momenta of the whole fully isolated system. Here we assume
that Eq. \ref{eq:13} is well determined under a proper contour $D$
in the $p_0$ -- complex plane. 
\\
\begin{definition}
\cite{Ch_PRA}
The three-dimensional Lorentz absolute momentum space $\mathbf{Q}_{rel,3}$
spanned by the momentum $\mathbf{q}$ is defined as a space in which the auxiliary function
$\tilde{F}(q^2)$
is embedded so that, for spacelike $p_\mu$
\begin{equation}
  \widetilde{F}(\mathbf{q}^2)=\widetilde{G}(p_\mu^2) = p_\mu'^2 = \mathbf{q}^2 \geq 0
  \label{eq:14}
\end{equation}
\end{definition}
Here $p_\mu^2$, $p_\mu'^2$ are representations of the 4-momentum in arbitrary frames $S, S'$.
Note that the above equality which defines internal space $\mathbf{Q}_{rel,3}$
does not distinguish any reference frame $S$ in $M_4$.
By definition $\widetilde{F}(\mathbf{q}^2)$
is implicitly Lorentz-form-invariant whereas $\widetilde{G}(p_\mu^2)$
is explicitly Lorentz-form-invariant. In this way, the directly
measurable functions $G$ implicitly  the determine $L$ absolute 3-dimensional space $Q_{rel,3}$.
In accordance with the quantum canonical symmetry $\tilde{F}(q)$
is interpreted as the momentum representation in $\mathbf{Q}_{rel}$
of the Hilbert vector $F$ 
space $|F\rangle$, $\widetilde{F}(\mathbf{q}^2)=\langle \mathbf{q} | F \rangle$ 
\begin{equation}
  \langle \mathbf{y} | F \rangle = F(\mathbf{y}^2) =
  (2\pi\hbar)^{-3} \int d^3 q \widetilde{F} (\mathbf{q}^2) \exp [i (\mathbf{q}\mathbf{y})/\hbar]
  \label{eq:15}
\end{equation}
is implicitly Lorentz-form-invariant quantity. Finally the quantum
internal relational space $\mathbf{Q}_{rel,3}$ spanned by the vector $\mathbf{y}$ 
is defined  as the three-dimensional $\mathbf{x}$ space. 

Note that the class of $|F\rangle$'s
is restricted according to the convergence of the corresponding integrals. In particular interval 
$|\mathbf{y}| = F(\mathbf{y}^2) = r$ in $\mathbf{Q}_{rel,3}$ has no representation in $M_4$
while the interval $x^2=G(x^2)$ in $M_4$ has no representation in $\mathbf{Q}_{rel,4}$.
However, all typical bound states and Yukawa-type potentials have both
internal and external representations in $\mathbf{Q}_{rel,4}$ and $M_4$.

Now, according to the quantum canonical symmetry $\mathbf{Q}$ in the Schrödinger representation 
$\hat{\mathbf{y}}=\mathbf{y}$, $\hat{\mathbf{q}}=-i\hbar \overline{\nabla}_y$
one can introduce an $L$-absolute internal Hamiltonian operator $\widehat{h}(\hat{\mathbf{y}}, \hat{\mathbf{q}})$
in the configuration-space $\mathbf{Q}_{rel,3}$, and the corresponding Schrödinger
equation of  the internal motion of an isolated system $S_M$ can be written in the form:
\begin{equation}
  i\hbar\frac{\partial}{\partial \tau} | \psi \rangle = \hat{h} | \psi \rangle
  \label{eq:16}
\end{equation}
where the internal Hamiltonian $\hat{h}$ is spanned by
$\hat{\mathbf{y}}$ and $\hat{\mathbf{q}}$ and, consistent with the relativistic kinematics, defines the absolute evolution parameter $\tau$
which completes the space $\mathbf{Q}_{rel,4}$
into the 4-dimensional continuum of the internal space-time $S_M$
and hence it is the Cartesian product $\mathbf{Q}_{rel,4}=\mathbf{Q}_{rel,3} \times T$
which operates with the two absolute intervals $r=|\mathbf{y}|$, $\Delta\tau=\Delta t$,
which coincide with the geometry intervals $\mathbf{Q}_{rel,4}$
Eq. \ref{eq:12}. Thus the nonrelativistic limit of $\mathbf{Q}_{rel,3}$ 
is the space $\mathbf{Q}_{rel,3}^G$
immersed in the Galilean space-time, while the internal time $\tau$
coincides up to an arbitrary translation constant with the $G$-absolute Newtonian time.

Chyliński argued that the space  $\mathbf{Q}_{rel,3}$
provides the first background of directly unobservable structures of
quantum microsystems, and  the   $L$-symmetry becomes a limiting symmetry of $\mathbf{Q}_{rel,3}$
conditioned by infinite inertia of interacting constituents
including measuring devices in the asymptotic zone of scattering. An
important novelty introduced by $\mathbf{Q}_{rel,3}$
is the nonlocality of the L-form invariant form factors $G(x^2)$,
which allows correlations over spatial intervals. Moreover, Chyliński
showed, that the concept of $\mathbf{Q}_{rel,4}$
space leads to potentially  measurable effects in elastic, inelastic 
and dissociation only in the processes in which composite particles
participate. Therefore, the quantum electrodynamics of point particles
is left unmodified. 

It is interesting that this
very rough model, ignoring the quark structure of the nucleon for
elastic form factor electron–nucleon collisions model predicts
proton mean-square radius 0.81fm \cite{ChPolA}, while recent experiments provide
values: 0.843fm \cite{Pohl}, 0.833fm \cite{Bezginov, Karr} 0.8414fm  
\cite{CODQTA}. These experimental discrepancies indicate that the proton
radius problem still remains open.

\section{Conclusion}

The concept of quantum information is central to the quantum model. In particular, it is shown that
``many fundamental features of quantum systems can be understood as consequences of the constraints that limited information content imposes on the systems'' \cite{Brukner}. However, this does not translate easily into the interpretational problems of quantum theory.
We tried to determine the status of quantum information and its place in quantum theory
within the QRS hypothesis.
In this aim we have introduced the correspondence principle between the physical structure represented by the set of observer-independence data, and the mathematical structure of the model that it describes. This principle allows us to interpret the wave function as a  mathematical representation (image) of quantum information that is relational in nature.
Consequently we have introduced an internal $Q_{rel}$ space which precedes the space of correlation events $Q_{cor}$.
We assumed that the mathematical structure of this hypothetical space is isomorphic to the tensorial Hilbert space structure and pointed out limitations of this assumption.
In this approach quantum correlations are interpreted as
limiting relations requiring decoherence and irreversibility
of recorded results.
Hence, the QRS hypothesis naturally excludes Wigner's type paradoxes.
We discussed some possible consequences of QRS hypothesis,
in particular, we showed that assumption that quantum
relations precede quantum correlations is compatible with
recent experimental and theoretical results \cite{Nandi,Time Delays}.

Finally, we consider Chyliński's quantum relational continuum space model, which leads to potentially observable effects in elastic, inelastic, and dissociation processes only for bound states. Unfortunately, Chyliński's articles went unnoticed, and to our knowledge, the
theoretical predictions have not been tested experimentally. Even
though 40 years have passed, the idea of Chyliński's internal
quantum space still seems interesting, especially in the context of
the quantum correlations phenomena.
In particular, it would be interesting to extend Chyliński's model
to include the spin degrees of freedom.

We believe that our approach, when properly developed, will lead to a deeper understanding of the role of quantum information in quantum theory.

\bmhead{Acknowledgments}
The author would like to especially thank L. Newelski for his suggestion to replace the isomorphic principle with
a correspondence principle that goes beyond a purely mathematical context. Special thanks are due to A. Wójcik for a panoramic, deep insight into the idea of this paper, and M. Eckstein and K. Horodecki for their deep and constructive criticism. The author would also like to thank M. Eckstein and P. Horodecki for a sequence of inspiring discussions and also M. Czachor, A. Grudka, A.
Miranowicz and M. Nowakowski for their discussion, and Ł. Pankowski for help with editing the manuscript. Finally, I would like to thank the anonymous Reviewers for their searching and inspiring criticism.

\bmhead{Funding}
The work was partly supported by the Foundation for Polish Science (IRAP Project, ICTQT, Contract No. MAB/2018/5, co-financed by the EU within the Smart Growth Operational Programme).


\begin{thebibliography}{999}

\bibitem{HHHH} Horodecki, R., Horodecki, P., Horodecki, M., Horodecki, K.: Quantum entanglement. Rev. Mod. Phys. \textbf{81}, 865 (2009)
\bibitem{Peres} Fuchs, C., Peres, A.: Quantum Theory Needs No `Interpretation'. Physics Today \textbf{53} (3), 70 (2000)
\bibitem{Bohm} Bohm, D.: A suggested interpretation of the quantum theory in terms of “hidden variables. I. Phys. Rev. \textbf{85}, 166–179 (1952)
\bibitem{Durr} Dürr, D., Goldstein, S., Zanghì, N.: Quantum physics without quantum philosophy. Springer. Heidelberg (2013)
\bibitem{Everett} Everett, H.: ‘Relative state’ formulation of quantum mechanics. Rev. Mod. Phys. \textbf{29}, 454 (1957)
\bibitem{Fuchs_RMP} Christopher Fuchs, A., Schack, R.: “Quantum-bayesian coherence”. Rev. Mod. Phys. \textbf{85}, 1693 (2013)
\bibitem{Rovelli} Rovelli, C.: Relational quantum mechanics. Int. J. Th. Phys. 35, 1637 (1996)
\bibitem{Mermin} Mermin, D.: What is quantum mechanics trying to tell us? American Journal of Physics \textbf{66}, 753 (1998)
\bibitem{Fayngold} Fayngold, M.: ,On the “Only Fields” interpretation of Quantum Mechanics arXiv:2111.07242
\bibitem{Vaidman} Vaidman, L.: “All is $\Psi$” Journal of Physics: Conference Series, \textbf{701}, 012020 (2016)
\bibitem{Vedral} Vedral, V.: The Everything-is-a-Quantum-Wave Interpretation of Quantum Physics. Quantum Rep. \textbf{5}, 475 (2023)
\bibitem{Deutsch} Deutsch, D.: Quantum theory, the Church–Turing principle and the universal quantum computer. Proc. R. Soc. Lond. A Math. Phys. Sci. \textbf{400}, 97 (1985).
\bibitem{Feynman} Feynman, R.P.: Quantum mechanical computers. Foundat. Phys. \textbf{16}, 507 (1986) 
\bibitem{Bennett_6} Bennett, C.H., Brassard, G., Crépeau, C., Jozsa, R., Peres, A. Wootters, W.K.: Teleporting an Unknown Quantum State via Dual Classical and Einstein-Podolsky-Rosen Channels. Phys. Rev. Lett. \textbf{70}, 1895 (1993)
\bibitem{Henderson_Vedral} Henderson, L., Vedral, V.: Classical, quantum and total correlations. J. Phys. A: Math. Theor. \textbf{34}, 6899 (2001)
\bibitem{Ollivier} Ollivier, H., Zurek, W. H.: Quantum Discord: A Measure of the Quantumness of Correlations. Phys. Rev. Lett. \textbf{88}, 017901 (2002)
\bibitem{OHH} Oppenheim, J., Horodecki, M., Horodecki, P., Horodecki, R.: Thermodynamical Approach to Quantifying Quantum Correlations. Phys. Rev. Lett. \textbf{89}, 180402 (2002)
\bibitem{Aspect}Aspect, A.:Closing the Door on Einstein
and Bohr’s Quantum Debate., Physics \textbf{8}, 123 (2015)
\bibitem{Raedt} De Raedt, H., Katsnelson, M. I., Jattana, M. S., Mehta, V., Willsch, M., Willsch, D., Michielsen, K., Jin, F.: Can foreign exchange rates violate Bell inequalities?, Annals of Physics \textbf{469}, 169742 (2024)
 \bibitem{Landsman} Landsman, N.P.: Between classical and quantum.  Handbook of the Philosophy of Science 2. 417-553 (2006)
\bibitem{Zwolak} Zwolak, M., Quan, H.T., Zurek, W.H.: Quantum Darwinism in a mixed environment. Phys. Rev. Lett. \textbf{103}, 110402 (2009)
\bibitem{RH_Korbicz} Horodecki, R., Korbicz, J.K., Horodecki, P.: Quantum origins of objectivity. Phys. Rev. A \textbf{91}, 032122 (2015) 
\bibitem{Paweł-Brandao-NC}  Brandão, F. G. S. L., Marco Piani, M., Horodecki, P.: Generic emergence of classical features in quantum Darwinism. Nat. Comm. \textbf{6}, 7908 (2015) 
\bibitem{Le} Le, T.P., Olaya-Castro, A.: Strong Quantum Darwinism and Strong Independence are equivalent to Spectrum Broadcast Structure. Phys. Rev. Lett. \textbf{122}, 010403 (2019)
\bibitem{Renner} Frauchige, D., Renner, R.: Quantum theory cannot consistently describe the use
of itself. Nat. Comm. \textbf{9}, 3711 (2018)
\bibitem{Żukowski} Żukowski, M., Markiewicz M.:  Physics and Metaphysics of Wigner's Friends: Even performed pre-measurements have no results.  Phys. Rev. Lett. \textbf{126}, 130402 (2021)  
\bibitem{Banach} Banach, S., Tarski, A.: Sur la 
d´ecomposition des ensembles de points en parties respectivement congruentes. Fundamenta Mathematicae \textbf{6}, 244–277 (1924)
\bibitem{Eckstein_P} Eckstein, M., Horodecki,P.: The Experiment Paradox in Physics, Foundations of Science  \textbf{27}, 1–15 (2022)
\bibitem{Raedt_EPR} De Raedt, H., Katsnelson, M. I., Jattana, M. S., Mehta, V., Willsch, M., Willsch, D., Michielsen, K., Jin, F.:
 Einstein-Podolsky-Rosen-Bohm experiments: a discrete data driven approach. Annals of Physics, \textbf{453}, 169314, (2023)
\bibitem{CODATA} Tiesinga, E., Mohr, P.J.,  Newell, D. B., Taylor, B. N.: CODATA recommended values of the fundamental physical constants: 2018*.Rev. Mod. Phys. \textbf{93}, 025010 (2021)
\bibitem{Stanford Encyclopedia of Philosophy} David, M.: The Correspondence Theory of Truth. First published Fri May 10, 2002; substantive revision Thu May 28, 2015.
\bibitem{Aristotle} Elsby, Ch.: Aristotle's Correspondence Theory of Truth and what does not exist. Logic and Logical 
\textbf{25}, 57 (2016)

\bibitem{Weinberg} Weinberg, S.: The trouble with Quantum Mechanics. (2017); available at: http://quantum.phys.unm.edu/466-17/Quantum Mechanics Weinberg.pdf
\bibitem{Stuckey} Stuckey, W., McDevitt, T., Silberstein, M.:  No Preferred Reference Frame at the Foundation of Quantum Mechanics. Entropy \textbf{24}, 12 (2022)
\bibitem{RH_comment}  Horodecki, R.: Comment on ‘Quantum principle of relativity’, New J. Phys. 25 128001 ( 2023)
\bibitem{Heisenberg} Heisenberg, W.: Niels Bohr and the Development of Physics. (Pergamon) (1955)
\bibitem{Private_Eckstein} Eckstein, M.: Private communication (email) 
\bibitem{Bethe_Salpeter} Betthe, H.A., Salpeter, E.E.: A Relativistic Equation for Bound-State Problems. Phys. Rev. \textbf{84},1231 (1951)
\bibitem{Ch_PRA} Chyliński, Z.: Internal symmetry of quantum systems and vertices of composite particles. Phys. Rev. A \textbf{32}, 764  (1985)
\bibitem{Klein} Klein, U.: Quantizing Galilean spacetime - A reconstruction of Maxwell's equations in empty space. arXiv:2304.11380
\bibitem{Chyliński_książka}  Chyliński, Z.: Kwanty a relatywistyka.Towarzystwo Autorów I Wydawców Prac Naukowych ‘Universytas’, Kraków 1992
\bibitem{Brukner} Brukner, C.: Information theoretic foundations of quantum theory (available at: www.iqoqi-vienna.at/research/brukner-group/information-theoretic-foundations-of-quantum-theory) (Accessed 11 November 2021)
\bibitem{Walls} Walls, S.M., Ford, I.J. Memory in a sequence of weak and short duration measurements of non-commuting observables. 	arXiv:2402.08737
\bibitem{Bargmann} Bargmann, V.: On Unitary Ray Representations of Continuous Groups, Annals of Mathematics. \textbf{59}, 1 (1954)
\bibitem{Newton_Wigner} Newton, T.D., Wigner, E.P.: Localized states for elementary systems. Rev. Mod. Phys. \textbf{21}, 400-408 (1949)
\bibitem{EPR} Einstein, A., Podolsky, B., Rosen, N.:     Can Quantum-Mechanical Description of Physical Reality Be Considered Complete? Phys. Rev. \textbf{47}, 777 (1935)
\bibitem{Schrödinger} Schrödinger, E.: Discussion of Probability Relations between Separated Systems. Math. Proc. Cambridge Philos. Soc. \textbf{31}, 555 (1935)
\bibitem{Jozsa} Jozsa, R.: Illustrating the concept of quantum information. IBM J. Res. Dev. \textbf{48}, 79 (2004)
\bibitem{Toth} Gühne, O., Tóth, G.: Entanglement detection. Phys. Rep. \textbf{74}, 1-75 (2009)
\bibitem{Nandi} Nandi, S. et al.: Generation of entanglement using a short-wavelength seeded free-electron laser. Sci. Adv. \textbf{10}, eado0668 (2024)
\bibitem{Time Delays}Jiang, W. C.,  et al.: 
Time Delays as Attosecond Probe of Interelectronic Coherence and Entanglement. Phys. Rev. Lett. \textbf{133}, 163201 (2024)
\bibitem{Kutak}Hentschinski, M., Kharzeev, D., Kutak, K., Tu, Z.: QCD evolution of entanglement entropy. arXiv:2408.01259 (Final version published in ROPP)
\bibitem{Bennett} Bennett C.H.: Quantum information. Phys. Scr. \textbf{1998}, 210 (1998)
\bibitem{Vedral_2}  Amico, L.,  Fazio, R., Osterloh, A., Vedral, V.: Entanglement in many-body systems,Rev. Mod. Phys. \textbf{80}, 517 (2008)
\bibitem{Zeilinger_RMP} Pan, J.-W., Chen, Z.-B., Lu, C.-Y., Weinfurter, H., Zeilinger, A., Żukowski, M.:  Multiphoton entanglement and interferometry. Rev. Mod. Phys. \textbf{84}, 777 (2012)
\bibitem{Modi} Modi, K., Brodutch, A., Cable, H., Paterek, T., Vedral, V.: The classical-quantum boundary for correlations: Discord and related measures. Rev. Mod. Phys. \textbf{84},1655 (2012)
\bibitem{Bennett_Brassard} Bennett, C. H., Brassard,  G.: Quantum Cryptography: Public Key Distribution and Coin Tossing. In Proceedings of the IEEE International Conference on Computers, Systems and Signal Processing, 175. IEEE Computer Society Press, New York, Bangalore, India (1984).
\bibitem{Ekert91} Ekert, A. K.: Quantum cryptography based
on Bell’s theorem. Phys. Rev. Lett. \textbf{67}, 661 (1991)
\bibitem{Deutsch_Jozsa} Deutsch, D., Jozsa, R.: Rapid solution of problems by quantum computation.
Proceedings of the Royal Society of London. Series A: Mathematical and Physical Sciences \textbf{439}, 553 (1992)
\bibitem{Shor} Shor, P. W.: Polynomial-Time Algorithms for Prime Factorization and Discrete Logarithms on a Quantum Computer. SIAM Journal on Computing \textbf{26}, 1484–1509 (1997)
\bibitem{Grover} Grover, L. K.: Quantum Mechanics Helps in
    Searching for a Needle in a Haystack. Phys. Rev. Lett. \textbf{79}, 325(1997)
\bibitem{Paweł_aktywacja} Horodecki, P., Horodecki, M., Horodecki, R.: Bound Entanglement Can Be Activated. Phys. Rev. Lett. \textbf{82}, 1056 (1999)
\bibitem{Paweł-Ekert} Horodecki, P., Ekert, A.: Method for Direct Detection of Quantum Entanglement.
Phys. Rev. Lett. \textbf{89}, 125003 (2002)
\bibitem{Michał_Nature} Horodecki, M., Oppenheim, J., Winter, A.: Quantum State Merging and Negative Information. Nature \textbf{436}, 673 (2005)
\bibitem{Bourennane} Nawareg, M., Muhammad, S., Horodecki, P., Bourennane, M.:  Science  Adv. \textbf{3}, e160248 (2017)
\bibitem{Karol_Winter} Yang, D., Horodecki, K., Winter, A.: Distributed Private Randomness Distillation. Phys. Rev. Lett. \textbf{123}, 170501 (2019)
\bibitem{Ingarden} Ingarden, R.S.: Quantum information theory. Rep. Math. Phys. \textbf{10}, 43 (1976)
\bibitem{Schumacher95} Schumacher, B.: Quantum coding. Phys. Rev. A \textbf{51}, 2738 (1995).
\bibitem{IBM} Horodecki, R., Horodecki, M., Horodecki, P.: Quantum information isomorphism: beyond the dillemma of Scylla of ontology and Charybdis of instrumentalism, IBM J. Res. Dev. \textbf{48}, 139 (2004)
\bibitem{Horodecki_QI} Horodecki,R.: Quantum information, Acta Phys. Pol. \textbf{139}, 197 (2021)
\bibitem{Zanardi}Zanardi, P.: Virtual Quantum Subsystems. Phys.Rev.Lett. \textbf{87}, 077901 (2001)
\bibitem{Dugic}Dugic, M., Jeknic-Dugic, J.: What is ”system”: the information-theoretic arguments. Int J Theor Phys. \textbf{47}, 805-813 (2008)
\bibitem{Dugic2}Jeknic-Dugic, J., Arsenijevic, M., Dugic, M.: "Quantum Structures: A View of the Quantum World", LAP Lambert Academic Publishing, Saarbrucken, (2013)
\bibitem{Dugic3} Kastner, R. E. Jeknic-Dugic, J., Jaroszkiewicz, G. Eds., “Quantum Structural Studies”, World Scientific, Singapore, (2017)
\bibitem{Vedral3} V. Vedral: Entanglement in the second quantization formalizm,
Central Eur. J. Phys. \textbf{1}, 289 (2003)
\bibitem{Caban_Podlaski} Caban, P., Podlaski, K., Rembieliński, J., Smoliński, K. A., Walczak, Z.,
Entanglement and tensor product decomposition for two fermions, J. Phys. A \textbf{38}, L79
\bibitem{Hulse_Schumacher} Hulse, A., Schumacher B.: Quantum meronomic frames. arXiv:1907.04899v1
\bibitem{Czachor_N} Czachor, M., Nowakowski, M.:  Relativity of spacetime ontology: When correlations in space become correlata in time. arXiv:2311.13879
\bibitem{Nagele} Nagele, Ch., Ilo-Okeke, E. O. P., Rohde, P., Dowling, J. P., Byrnes, T.: Relativity of quantum states in entanglement swapping. Physics Letters A, \textbf{384}, 126301 (2020)
\bibitem{Caban}Paweł Caban, P., Hiesmayr, B.C.: Is bound entanglement Lorentz invariant?, Scientific Reports \textbf{13}, 11189 (2023)
\bibitem{Banaszek} Banaszek, K., Kunz, L., Jachura, M., Jarzyna, M.: : Quantum Limits in Optical Communications. J. Lightwave Technol. \textbf{38}, 2741 (2020)
\bibitem{Hagg} Haag, R.: "Quantum field theories with composite particles and asymptotic conditions". Physical Review. \textbf{112}, 669–673 (1958)
\bibitem{Peskin} Peskin, M.E., Schroeder, D.V.: An Introduction to Quantum Field Theory. Addison-Wesley, Reading (1995)
\bibitem{Di Biagio} Di Biagio, A., Howl, R., Brukner, C., Rovelli, C., Christodoulou, M.: Relativistic locality can imply subsystem locality: arXiv:2305.05645
\bibitem{Gisin_QT}  Gisin, N.: Quantum correlations in Newtonian space and time: arbitrarily fast communication or nonlocality. In: Struppa D., Tollaksen J. (eds) Quantum Theory: A Two-Time Success Story. Springer, Milano. (2014) https://doi.org/10.1007/978-88-470-5217-8 
\bibitem{Clauser_Shimony} Clauser, J.F., Shimony, A.: Bell's theorem. Experimental tests and implications. Rep. Prog. Phys. \textbf{41}, 1881 (1978)
\bibitem{Bohr} Bohr, N.: Atomic Theory and the Description of Nature. Univ. Press Cambridge (1934)
\bibitem{Clauser}Clauser, J. F.: Laboratory-Space and Configuration-Space Formulations of Quantum Mechanics, Versus Bell–Clauser–Horne–Shimony Local Realism, Versus Born’s Ambiguity, In book: Quantum Arrangements, Contributions in Honor of Michael Horne (pp.35-91)  (2021) https://doi.org/10.1007/978-3-030-77367-0
\bibitem{Gisin_Sunday} Gisin, N.: Sundays in a quantum engineer’s life. chapter 13, In: Bertlmann R, Zeilinger A (eds) Quantum [un]speakables, from Bell to quantum information. Proceedings of the 1st quantum
[un]speakables conference. Springer International, Switzerland  (2002)
\bibitem{Clauser_Gdansk} Clauser, J.: Ingarden Memorial Lecture "Experimental proof that quantum entanglement is real", National Quantum Information Centre in Poland ,V R.S. Ingarden Memorial Session  November 27. (2024)
\bibitem{Esfeld} Esfeld, M.: Quantum entanglement and a metaphysics of relations. Studies in History and Philosophy of Science Part B: Studies in History and Philosophy of Modern Physics.  \textbf{35},  601, (2004)
\bibitem{Mahdavipour} Mahdavipour, K., Nosrati, F., Sciara, S., Morandotti, R., Franco, R.L.: Generation of genuine multipartite entangled states via indistinguishability of identical particles. arXiv:2403.17171
\bibitem{Pranzini}Pranzini, N.,  Verrucchi, P.: On the reliability and accessibility of quantum measurement apparatuses. Phys. Rev. A \textbf{109}, 032203 (2024)
\bibitem{Gallego}Gallego, M.: Hilbert space separability and the Einstein-Podolsky-Rosen state. arXiv:2412.01897
\bibitem{Weaver}Weaver, N.: Quantum Graphs as Quantum Relations.. The Journal of Geometric Analysis. https://doi.org/10.1007/s12220-020-00578-w
\bibitem{Halvorson}Halvorson, H.: Complementarity of representations in
quantum mechanics. Stud. Hist. Philos. Mod. Phys. \textbf{35}, 45 (2004)
\bibitem{Kastner2021} Kastner, R.: Unitary Interactions do not yield outcomes: Attempting to model ‘Wigner’s Friend’. Found. of Phys. \textbf{51},1 (2021)
\bibitem{Sewell} Sewell, G.L.:On the mathematical  of the quantum measurement problem. Rep. Math. Phys. \textbf{56}, 271-290 (2005)
\bibitem{Sen} Sen, R.N.: Homer nodded once more. Von Neumann's misreading of the Compton-Simon experiment and its fallout. arXiv:2302.14610
\bibitem{Zurek} Ollivier, H., Poulin, D., Zurek, W. H.: Objective Properties from Subjective Quantum States: Environment as a Witness. Phys. Rev. Lett. \textbf{93}, 220401 (2004)
\bibitem{Gorini} Gorini, V., Kossakowski, A., Sudarshan, E. C. G.: Completely positive dynamical semigroups of
N-level systems. J. Math. Phys. \textbf{17}, 821 (1976)
\bibitem{Lindblad} Lindblad,G.: On the Generators of Quantum Dynamical Semigroups. Comm. Math. Phys. \textbf{48},
119–130 (1976)
\bibitem{Brukner_noRowelli} Brukner, C.: Qubits are not observers -- a no-go theorem. arXiv:2107.03513 
\bibitem{Jacques L. Pienaar} Pienaar,  J. L.: A quintet of quandaries: five no-go theorems for Relational Quantum Mechanics. Found Phys \textbf{51}, 97 (2021)

\bibitem{Weinert} Weinert, F.: ‘Minkowski Spacetime and Thermodynamics’, in V. Petkov (Ed.), Space, Time and Spacetime. Heidelberg, Berlin, New York: Springer (2010), pp. 239-56 (Fundamental Theories of Physics Vol. 167)
\bibitem{J_Geom_Phys} Miller, T., Eckstein, M., Horodecki, P., Horodecki, R.: Generally covariant N-particle dynamics. J. Geom. Phys. \textbf{160}, 103990 (2021).
\bibitem{Heller} Eckstein, M., Heller, M.:  Under review in Studies in History
and Philosophy of Science. arXiv:2202.07302 (2022).
\bibitem{Reichenbach} Hitchcock, C., Redei, M.: Reichenbach’s Common Cause Principle. In: The Stanford Encyclopedia of Philosophy. Ed. by Edward N. Zalta. Summer 2021. Metaphysics Research Lab, Stanford University  (2021)
\bibitem{Freericks} Freericks, J.K.: How to measure the momentum of single quanta. EPJ ST \textbf{20}, 3285 (2023)
\bibitem{ChPolB} Chyliński, Z.: Relationism of quantum physics. Acta Phys. Pol. B, \textbf{26}, 1547 (1995)
\bibitem{ChPolA} Chyliński, Z., Acta Phys. Pol. A \textbf{65}, 369 (1984).
\bibitem{Pohl} Pohl R, Antognini A, Nez F, Amaro FD, Biraben F, et al. (July 2010). "The size of the proton" (PDF). Nature. \textbf{466} (7303): 213–216.
\bibitem{Bezginov}  Bezginov, N.; Valdez, T.; Horbatsch, M.; Marsman, A.; Vutha, A. C.; Hessels, E. A. A measurement of the atomic hydrogen Lamb shift and the proton charge radius. Science. \textbf{365} 1007–1012
\bibitem{Karr} Karr, J.-P., Marchand, D., Voutier, E.: The proton size. Nature Reviews Physics. \textbf{2} 601–614 (2020)
\bibitem{CODQTA} Tiesinga, E., Mohr, P. J., Newell,D. B., Taylor, B. N.: CODATA recommended values of the physical Constants: 2018.J. Phys. Chem. Ref. Data \textbf{50}, 033105 (2021)

\end{thebibliography}
\end{document}